\documentclass[graybox]{svmult}

\usepackage{mathptmx}       
\usepackage{helvet}         
\usepackage{courier}        
\usepackage{type1cm}        
\usepackage{makeidx}         
\usepackage{graphicx}        
\usepackage{multicol}        
\usepackage[bottom]{footmisc}

\makeindex             

\begin{document}

\title*{Resistive MHD jet simulations with large resistivity}
\author{Miljenko \v{C}emelji\'{c}, Jose Gracia, Nektarios Vlahakis and
Kanaris Tsinganos}
\institute{Miljenko \v{C}emelji\'{c} \at 
TIARA, Academia Sinica,
 National Tsing Hua University, No. 101, Sec. 2,
Kuang Fu Rd., Hsinchu 30013, Taiwan \email{miki@tiara.sinica.edu.tw} \and  
Jose Gracia \at School of Cosmic Physics, Dublin Institute for Advanced    
Studies, 31 Fitzwilliam Place, Dublin 4, Ireland \email{jgracia@cp.dias.ie}
\and Nektarios Vlahakis \& Kanaris Tsinganos \at IASA and Section of
Astrophysics, Astronomy and Mechanics, Dpt. of Physics, Univ. of
Athens, Panepistemiopolis 15784 Zografos, Athens, Greece
\email{vlahakis,tsingan@phys.uoa.gr}}
%
%
\maketitle

\abstract*{Axisymmetric resistive MHD simulations for radially 
self-similar initial conditions are performed, using the NIRVANA code.
The magnetic diffusivity could occur in outflows above an
accretion disk\index{accretion disk}, being transferred from the underlying disk into the disk
corona by MHD\index{MHD} turbulence (anomalous turbulent diffusivity),
or as a result of ambipolar diffusion in partially ionized flows. We
introduce, in addition to the classical magnetic Reynolds number Rm, which
measures the importance of resistive effects in the induction equation, a
new number Rb, which measures the importance of the
resistive effects in the energy equation. We find two distinct regimes of
solutions in our simulations. One is the low-resistivity
regime, in which results do not differ much from ideal-MHD\index{MHD} solutions. In the
high-resistivity regime, results seem to show some periodicity in
time-evolution, and depart significantly
from the ideal-MHD\index{MHD} case. Whether this departure is caused by numerical or
physical reasons is of considerable interest for numerical
simulations\index{numerical simulations} and
theory of astrophysical outflows and is currently investigated.
}
\abstract{
Axisymmetric resistive MHD\index{MHD} simulations for radially   
self-similar initial conditions are performed, using the NIRVANA code.
The magnetic diffusivity could occur in outflows above an
accretion disk\index{accretion disk}, being transferred from the underlying disk into the disk
corona by MHD\index{MHD} turbulence (anomalous turbulent diffusivity),
or as a result of ambipolar diffusion in partially ionized flows. We
introduce, in addition to the classical magnetic Reynolds number Rm, which            
measures the importance of resistive effects in the induction equation, a
new number Rb, which measures the importance of the    
resistive effects in the energy equation. We find two distinct regimes of
solutions in our simulations. One is the low-resistivity
regime, in which results do not differ much from ideal-MHD\index{MHD} solutions. In the
high-resistivity regime, results seem to show some periodicity in  
time-evolution, and depart significantly
from the ideal-MHD\index{MHD} case. Whether this departure is caused by numerical or
physical reasons is of considerable interest for numerical
simulations\index{numerical simulations} and
theory of astrophysical outflows and is currently investigated.}
\section{Introduction}
\label{sec:1}
\begin{figure}[b]
\includegraphics[scale=.42]{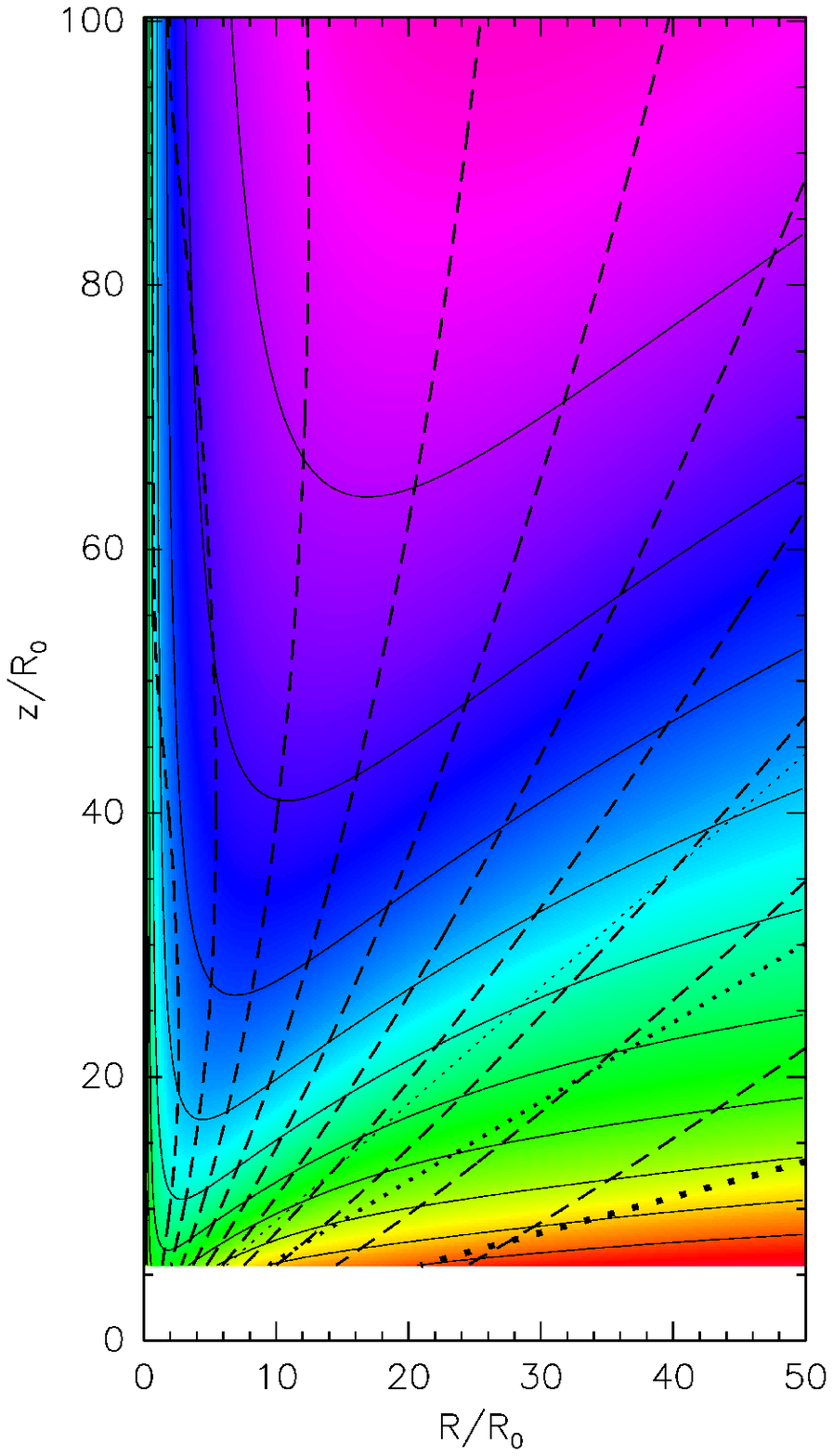}
\includegraphics[scale=.35]{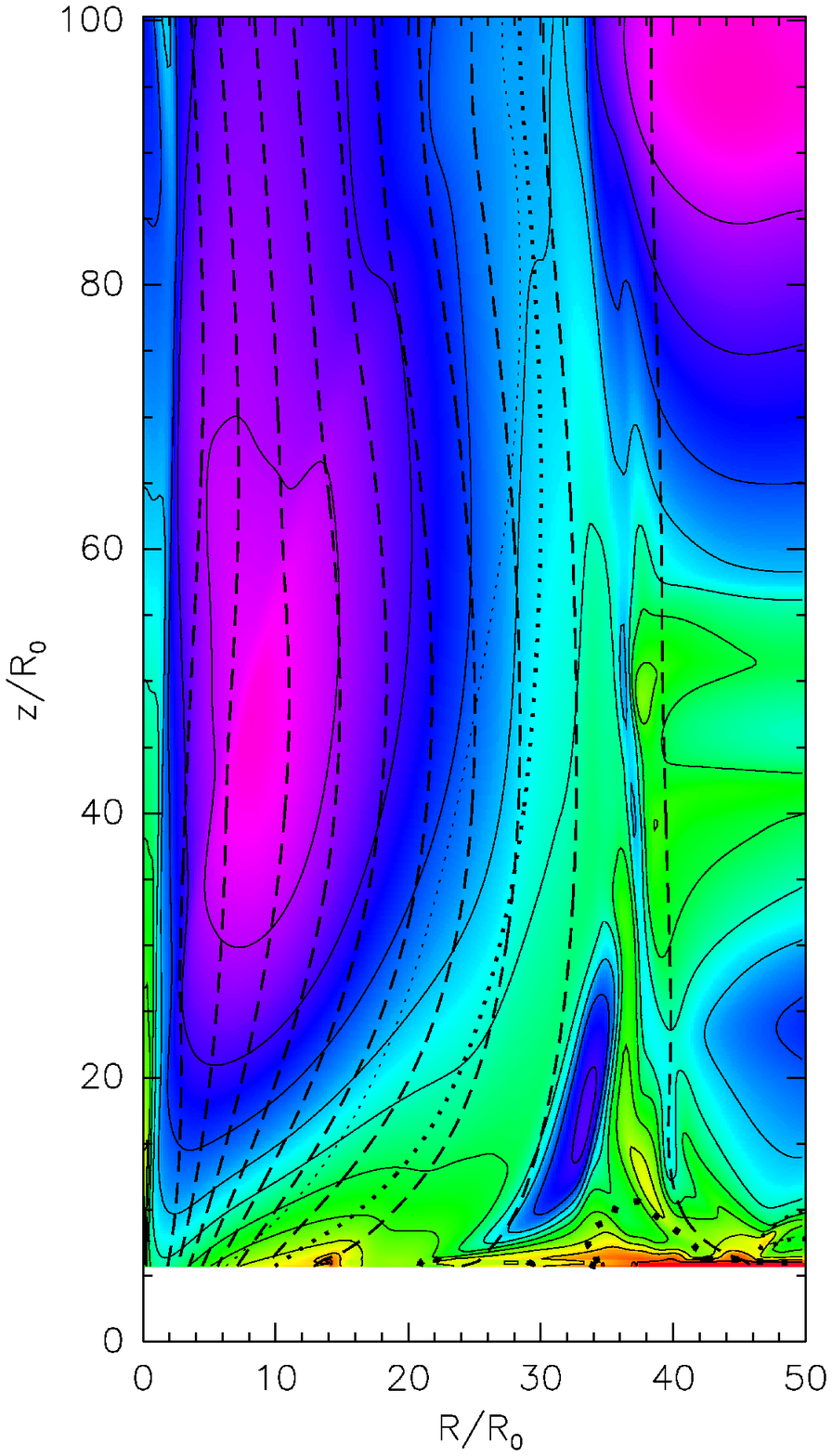}
\caption{The initial setup, which is slightly modified analytical solution,
is shown in the {\em Left} panel. The solid
lines represent logarithmically spaced isocontours of density. It is also
shown in colour grading, in red to violet colour, for the logarithm of 
density -1 to -4, respectively. In the {\em Right} panel shown is, in the same
grading, the
solution with large magnetic diffusivity. It
does not reach stationary state, and shows some periodicity in time
evolution. The dashed lines depict poloidal magnetic field lines,
and the dotted lines depict the fast magnetosonic, Alfven and slow magnetosonic
critical surfaces, top to bottom, respectively.
}
\label{fig1}
\end{figure}
 In Vlahakis \& Tsinganos (1998) general classes of self-consistent
ideal-MHD\index{MHD}
solutions have been constructed. In Vlahakis et al. (2000) Blandford \& Payne
(1982) model was analysed, and the problem with the terminal wind solution
(which was not  causally disconnected from the disk) has been solved. The
common deficiency of all radially self-similar models, a cut-off of the
solution at small cylindrical radii and also at some finite height above the
disk because of a strong Lorentz force close to the system's axis has been
corrected numerically. A search in the numerical simulations\index{numerical
simulations} for solutions
at larger distances from the disk has been performed in Gracia et al. (2006)
with NIRVANA code (version 2.0, Ziegler, 1998), and similar results were
obtained also using the PLUTO code in Matsakos et
al. (2008). Extension in the resistive-MHD\index{MHD} has been investigated in
\v{C}emelji\'{c}
et al. (2008) using the NIRVANA code and some of results we present here.
Results of these investigations could also have implications in the
numerical simulations\index{numerical simulations} of magnetospheric interaction in vicinity of
the young stellar objects, where the resistivity plays important
role.
\begin{figure}[b]
\includegraphics[scale=.78]{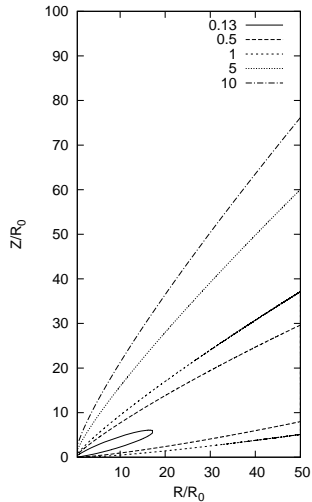}
\caption{
Value of  
$\beta/2 (VR/V_0R_0)$ for the analytical solution used as initial setup
here. This quantity gives the critical value of magnetic diffusivity $\eta$
 that corresponds to Rb=1.
}
\label{fig2}
\end{figure}

Our numerical simulations\index{numerical simulations} are initiated by the slightly modified analytical solutions
for radially self-similar flow from Vlahakis et al. (2000), and then 
evolved in the resistive MHD\index{MHD} simulations by NIRVANA code. Our initial setup
is shown in the left panel of Fig.\ref{fig1}.
\section{New characteristic number}
\label{sec:2}
In addition to the magnetic Reynolds number Rm=VR/$\eta$, which describes influence of
the magnetic diffusivity $\eta$ in the induction equation, we introduced a
new number, which describes the influence in the energy transport
equation-see Fig.\ref{fig2}. It can be written 
in terms of Rm and plasma beta
as Rb=Rm$\beta$/2. It is the ratio of the pressure term over the Joule heating
term in the energy equation. When Rb is
smaller or close to unity, which can happen even when Rm is much larger than
unity, the energy dissipation becomes important. It might define one additional
mode of resistive-MHD\index{MHD} solutions, indicated in our search for eventual onset of
super-critical resistive regime.

\section{Results}
\label{sec:3}
\begin{figure}[b]
\includegraphics[scale=.5]{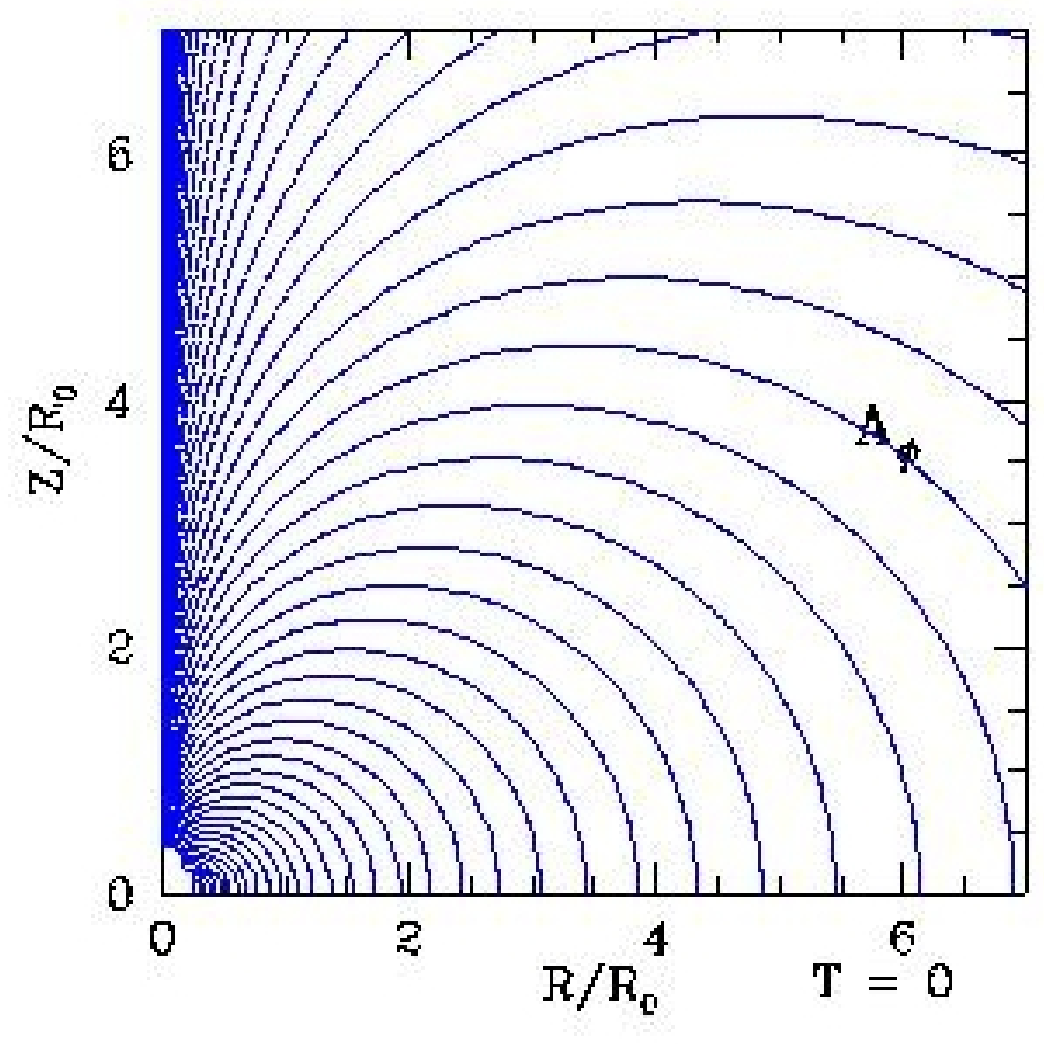}
\includegraphics[scale=.5]{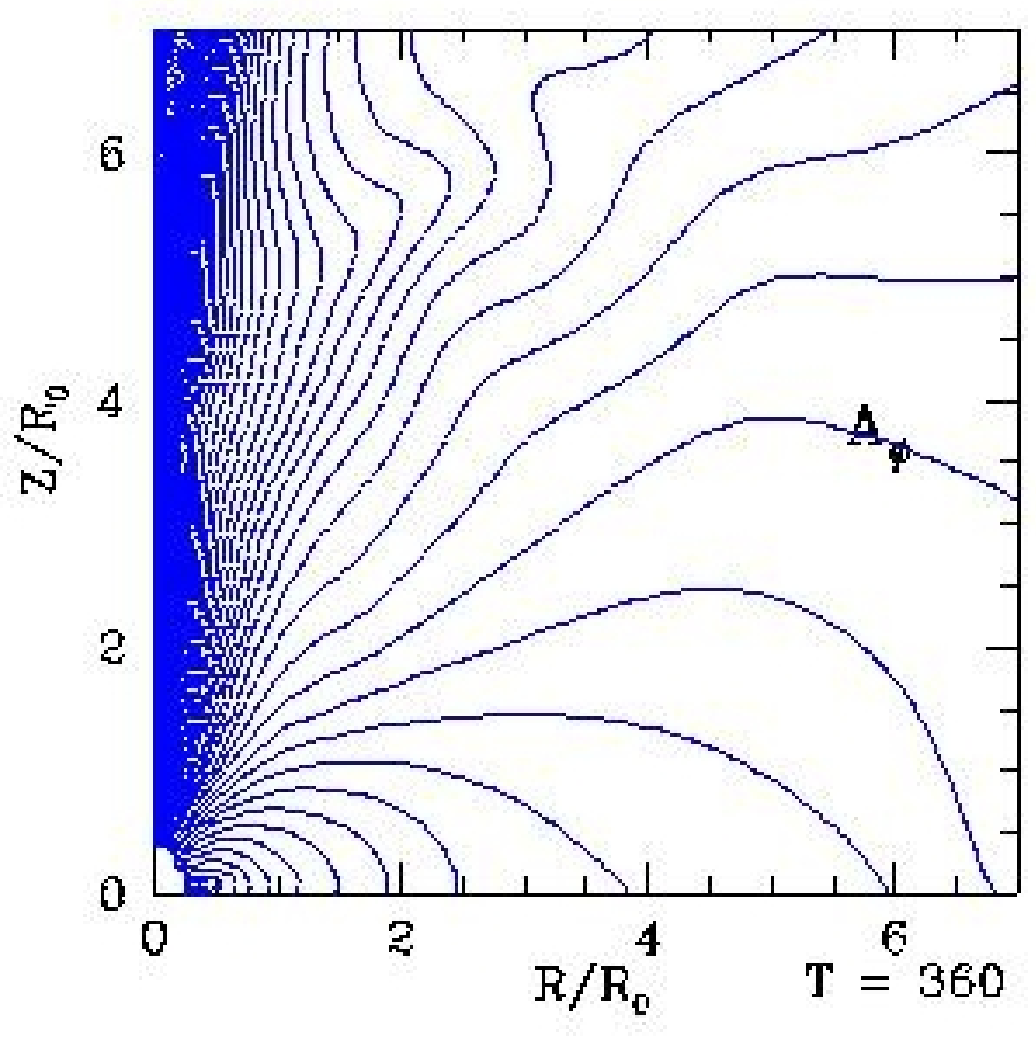}\\
\includegraphics[scale=.5]{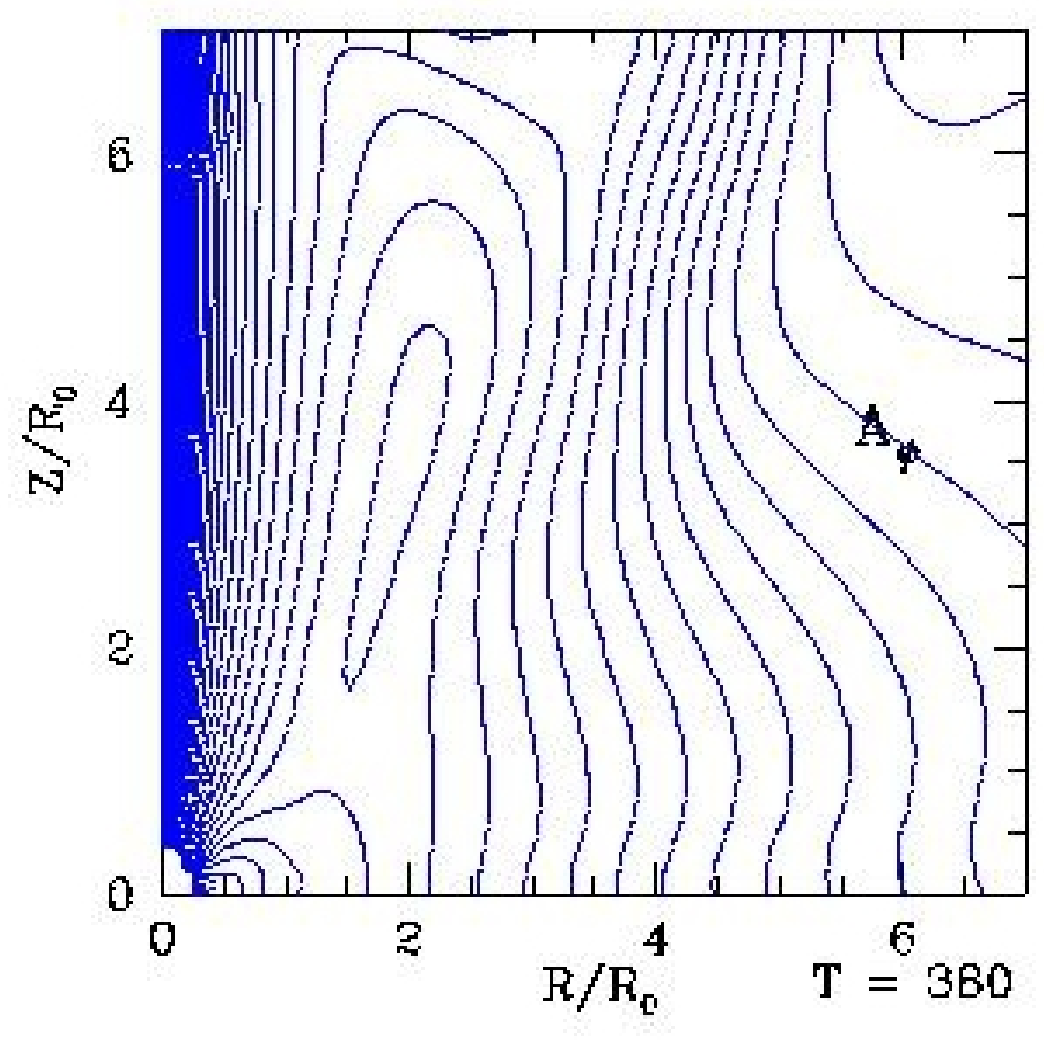}
\includegraphics[scale=.5]{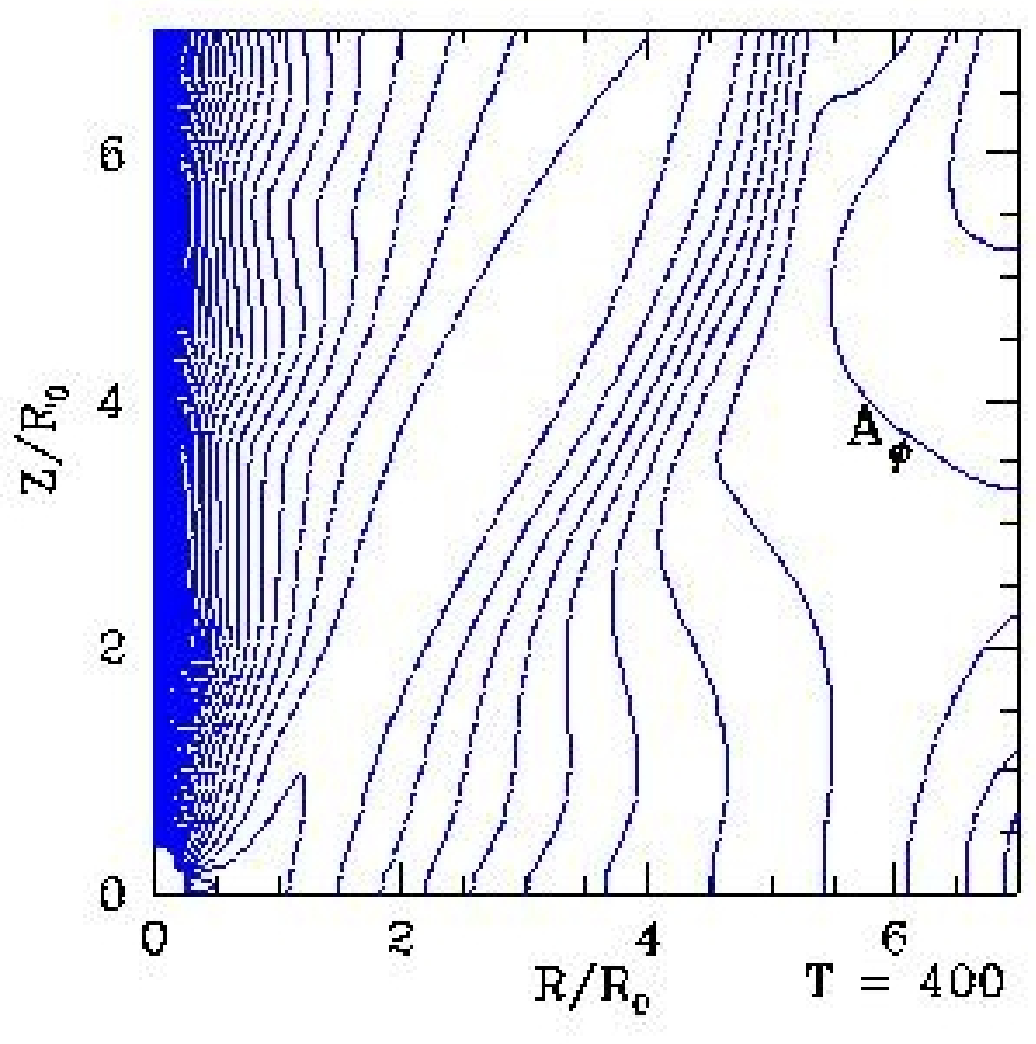}               
\caption{Reconnection and re-shaping of the magnetic field in the vicinity
of the young stellar object. Initially pure stellar dipole field reshapes
into the stellar and disk open field during the time-evolution. The time
 is measured in the number of rotations at the inner disk radius, which
is at R$_i$=3.0 in these simulations. Without the substantial resistivity,
reconnection does not occur and simulations terminate because of numerical
reasons.
}
\label{fig3}
\end{figure}
The resistive MHD\index{MHD} jets\index{jets} are similar to ideal-MHD\index{MHD} solutions for a finite
range of magnetic diffusivity, in which they reach a well defined stationary
state. This state only slightly differs from the initial state, as expected,
since the initial setup was slightly modified analytical 
stationary solution. Departure from the ideal-MHD\index{MHD} regime occurs for 
larger values of
magnetic diffusivity, above some critical value. One such result is shown in
the right panel of Fig.\ref{fig1}. We note possible existence of
the distinct super-critical regime in magnetic diffusivity in our simulation
for the outflows
initialised with a self-similar analytical solutions. We also define the new
characteristic number, which describes the influence of the resistivity on
the energy transport equation.
Physical parameters and the eventual periodicity of the super-critical resistive
solutions are currently under investigation. Such solutions might be
interesting for investigations of accretion flows in the vicinity of young
stellar objects, where the magnetic resistivity seems to play important role.

In the Fig.\ref{fig3} we show one case in numerical
simulations\index{numerical simulations} of
magnetospheric interaction in the closest vicinity of young stellar object
(\v{C}emelji\'{c} et al., 2009). These simulations have been performed with code ZEUS347,
which is our resistive version of Zeus-3D code (Fendt \& \v{C}emelji\'{c},
2002). Magnetic reconnection 
shows to play essential role
in re-shaping the initial stellar dipole, which enables the launching of outflows.
In our numerical simulations\index{numerical simulations}, if the resistivity is too small, reconnection
will not occur. Therefore, we need to investigate parameter space for
resistivity, and we need to understand what are the effects of large, and
not only negligible or very small resistivity.
\begin{acknowledgement}
This work was supported in part by EC's Marie Curie Actions-Human Resource
and Mobility within the JETSET network under contract MRTN-CT-2004005592.
M\v{C} expresses gratitude to TIARA/ASIAA in Taiwan for possibility to
use their Linux clusters and JETSET for supporting this collaboration.
\end{acknowledgement}
%
%
%

\begin{thebibliography}{99.}%
%
\bibitem{phys-journal} Blandford R. D., Payne D. G., 1982, 
MNRAS, 199, 883 
\bibitem{phys-journal} \v{C}emelji\'{c} M., Hsien S., Chiang T.-Y., 2009, in preparation
\bibitem{phys-journal} \v{C}emelji\'{c} M., Gracia J., Vlahakis N., Tsinanos K.,
2008, MNRAS, 389, 1022, 1032 
\bibitem{phys-journal} Fendt Ch., \v{C}emelji\'{c} M., 2002, A\&A, 395, 1045 
\bibitem{phys-journal} Gracia J., Vlahakis N., Tsinganos K., 2006,
MNRAS, 367, 201 
\bibitem{phys-journal} Matsakos T., Tsinganos K., Vlahakis N., Massaglia S.,
Trussoni E., 2008, A\&A, 477, 521 
\bibitem{phys-journal} Vlahakis N., Tsinganos K., 1998,
MNRAS, 298, 777 
\bibitem{phys-journal} Vlahakis N., Tsinganos K., Sauty C., Trussoni E.,
2000, MNRAS, 318, 417 
\bibitem{phys-journal} Ziegler U., 1998,
Comput. Phys. Commun., 109, 111

%
\end{thebibliography}
%

\end{document}